\documentclass[superscriptaddress,amsmath,amssymb,aps,prl,preprint,tightenlines]{revtex4-1}

\usepackage{graphicx,dcolumn,braket,bm}

\usepackage[T1]{fontenc} % if needed

\begin{document}

\preprint{RESCEU-7/19}

\title{Dark matter and baryon-number generation in quintessential inflation via hierarchical right-handed neutrinos}

\author{Soichiro Hashiba}
	\email{sou16.hashiba@resceu.s.u-tokyo.ac.jp}
	\affiliation{Department of Physics, Graduate School of Science, The University of Tokyo, Tokyo 113-0033, Japan}
	\affiliation{Research Center for the Early Universe (RESCEU), Graduate School of Science, The University of Tokyo, Tokyo 113-0033, Japan}
\author{Jun'ichi Yokoyama}
	\email{yokoyama@resceu.s.u-tokyo.ac.jp}
	\affiliation{Department of Physics, Graduate School of Science, The University of Tokyo, Tokyo 113-0033, Japan}
	\affiliation{Research Center for the Early Universe (RESCEU), Graduate School of Science, The University of Tokyo, Tokyo 113-0033, Japan}
	\affiliation{Kavli Institute for the Physics and Mathematics of the Universe (Kavli IPMU), WPI, UTIAS, The University of Tokyo, 5-1-5 Kashiwanoha, Kashiwa 277-8583, Japan}

\date{\today}

\begin{abstract}
Incorporating three generations of right-handed Majorana neutrinos to quintessential inflation, we construct a model which simultaneously explains inflation, dark energy, dark matter and baryogenesis.  These neutrinos have hierarchical masses $M_3 \sim 10^{13}\ {\rm GeV}, M_2 \sim 10^{11}\ {\rm GeV}, M_1 \sim 10\ {\rm keV}$ and are produced by gravitational particle production in the kination regime after inflation.  The heaviest, the intermediate, and the lightest account for reheating, CP violation of leptogenesis, and dark matter, respectively.  This model can be tested in various ways with forthcoming observations.

\if{We construct a model, which simultaneously explains inflation, dark energy dark matter and baryogenesis with Majorana right-handed neutrinos. We consider three generations of right-handed neutrinos with hierarchal masses $M_3 \sim 10^{13}\ {\rm GeV}, M_2 \sim 10^{11}\ {\rm GeV}$ and $M_1 \sim 10\ {\rm keV}$, non-minimal coupling between the scalar curvature $R$ and fermion $\psi$ such as $\frac{R}{\mu}\bar{\psi}\psi$ and a cutoff scale $\mu \sim 10^{15}\ {\rm GeV}$ in quintessential inflationary model. The heaviest, the intermediate and the lightest right-handed neutrinos explain the reheating process, the baryogenesis via the leptogenesis and the dark matter, respectively.}\fi

\end{abstract}

\maketitle

\textit{Introduction.}---Inflation in the early universe can solve a number of fundamental problems associated with the geometrical structure of the universe such as the horizon and flatness problems as well as the origin of curvature perturbation (for a review of inflation, see {\it e.g.} \cite{Sato:2015dga}). On the other hand, there remains many problems about the material contents of the universe, namely, the origin of baryon asymmetry, the origin and identities of dark matter, dark energy, which drives current accelerated expansion, and a concrete mechanism of inflation. In this letter, we work out these remaining problems by incorporating three generations of right-handed Majorana neutrinos in quintessential inflation \cite{Peebles:1998qn}. Our model not only realizes accelerated expansion of current and early universe in a single setup, but also explains the reheating process, baryogenesis and the origin of the cold dark matter simultaneously by gravitational production of these right-handed neutrinos, which are originally introduced in order to explain the small non-zero neutrino masses \cite{Yanagida:1979as,GellMann:1980vs} as indicated by neutrino oscillation experiments \cite{Fukuda:1998mi}. We use the natural units $c = \hbar = 8\pi G = 1$ throughout the letter.

\textit{Gravitational particle creation.}---Although the original model of quintessential inflation \cite{Peebles:1998qn} has already been ruled out by the recent observation of the cosmic microwave background (CMB) \cite{Akrami:2018odb}, there are several quintessential models consistent with observational data, such as those based on $\alpha$-attractors \cite{Dimopoulos:2017zvq,Akrami:2017cir}, which are also theoretically well motivated, or those using exponential potentials \cite{Hossain:2014zma,Geng:2017mic}. In these models, inflation is followed by a kinetic energy dominant era which is called kination and the transition from inflation to kination takes about one Hubble time. In this class of models, reheating is achieved by the gravitational particle creation \cite{Parker:1969au,Zeldovich:1971mw} induced by time variation of the metric from de Sitter to power-law cosmic expansion \cite{Kunimitsu:2012xx,Hashiba:2018iff}.

\begin{figure}[tbp]
\centering
\includegraphics[width=.47\textwidth]{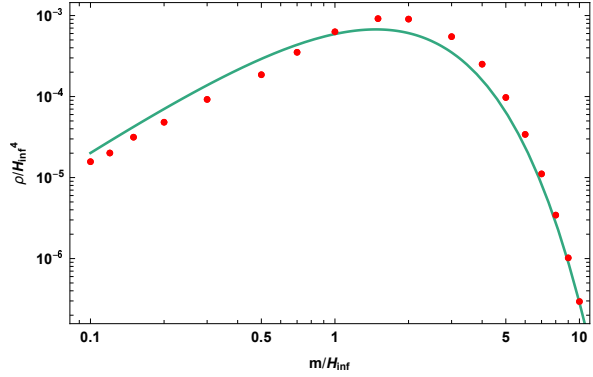}
\caption{\label{fig:rho} The energy density of gravitationally produced massive fermion. The red dots and the green line show the numerical result of (\ref{eommix}) and its fitting curve, respectively. In this graph, we set $\Delta t = 0.3 H_{\rm inf}^{-1}.$}
\end{figure}

Let us consider a massive fermion represented by a Dirac spinor field $\psi$ minimally (conformally) coupled to gravity in a spatially flat Friedmann-Lema\^{i}tre-Robertson-Walker (FLRW) metric, $ds^2 = g_{\mu\nu} dx^\mu dx^\nu = a^2(\eta) (- d\eta^2 + d\bm{x}^2)$, where $a(\eta)$ and $\eta$ denote the scale factor and the conformal time which satisfies $a\,d\eta=dt$, respectively. The Lagrangian of $\psi$ is
\begin{equation}
	\mathcal{L}_\psi = \sqrt{-g} \left( i\bar{\psi} \gamma^\mu(x) \nabla_\mu \psi - m\bar{\psi}\psi \right). \label{Lpsi1}
\end{equation}
Here, $\gamma^\mu(x)$ is the gamma matrices which satisfy
\begin{equation}
	\{\gamma^\mu(x),\gamma^\nu(x)\}=-2g^{\mu\nu}(x)I_4,
\end{equation}
$\bar{\psi} \equiv \psi^\dagger \gamma^0(x)$ is the Dirac adjoint of $\psi$ and $\nabla_\mu \equiv \partial_\mu - \Gamma_\mu$ are the covariant derivatives, where the spin connections $\Gamma_\mu(x)$ and the Christoffel symbols $\Gamma^\sigma_{\mu\nu}(x)$ satisfy
\begin{equation}
	[\Gamma_\nu(x),\gamma^\mu(x)] = \partial_\nu \gamma^\mu(x) + \Gamma^\mu_{\nu\rho}(x)\gamma^\rho(x).
\end{equation}
In a spatially flat FLRW metric, the gamma matrices are given by $\gamma^\mu(\eta) = a^{-1}(\eta)\gamma^\mu$, where and hereafter $\gamma^\mu$ with no argument denotes the gamma matrices in the Minkowski space. Hence, (\ref{Lpsi1}) becomes
\begin{equation}
	\mathcal{L}_\Psi = i\bar{\Psi} \gamma^\mu \partial_\mu \Psi - a m\bar{\Psi} \Psi, \label{Lpsi}
\end{equation}
where $\Psi \equiv a^{3/2}\psi$. Taking the variation of its action with respect to $\bar{\Psi}$, the principle of least action gives the Dirac equation
\begin{equation}
	\left( -i\gamma^\mu \partial_\mu + a m \right) \Psi = 0. \label{diraceq}
\end{equation}
This equation can be rewritten in terms of the spinor mode functions $u_{A, B} (\bm{k}, \eta)$ as
\begin{equation}
	i \partial_\eta \left( \begin{array}{c} u_A \\ u_B \end{array} \right) = \left( \begin{array}{cc} a m & k \\ k & -a m \end{array} \right) \left( \begin{array}{c} u_A \\ u_B \end{array} \right).
\end{equation}
We can separate a mixing term by inserting a pair of unitary matrices and obtain
\begin{widetext}
\begin{equation}
	i \partial_\eta \left( \begin{array}{c} \tilde{u}_A \\ \tilde{u}_B \end{array} \right) = \left( \begin{array}{cc} \omega_{\rm eff} & 0 \\ 0 & -\omega_{\rm eff} \end{array} \right) \left( \begin{array}{c} \tilde{u}_A \\ \tilde{u}_B \end{array} \right) + \left( \begin{array}{cc} 0 & \frac{am H k}{2\omega_{\rm eff}^2}  \\ -\frac{am H k}{2\omega_{\rm eff}^2} & 0 \end{array} \right) \left( \begin{array}{c} \tilde{u}_A \\ \tilde{u}_B \end{array} \right), \label{eommix}
\end{equation}
\end{widetext}
where
\begin{equation}
	\left( \begin{array}{c} \tilde{u}_A \\ \tilde{u}_B \end{array} \right) = \left( \begin{array}{cc} \frac{k}{\sqrt{k^2 + (am - \omega_{\rm eff})^2}} & \frac{-am + \omega_{\rm eff}}{\sqrt{k^2 + (am - \omega_{\rm eff})^2}} \\ \frac{k}{\sqrt{k^2 + (am + \omega_{\rm eff})^2}} & \frac{-am - \omega_{\rm eff}}{\sqrt{k^2 + (am + \omega_{\rm eff})^2}} \end{array} \right) \left( \begin{array}{c} u_A \\ u_B \end{array} \right),
\end{equation}
$\omega_{\rm eff} \equiv \sqrt{k^2 + a^2 m^2}$ and $H$ is the Hubble parameter \cite{Chung:2011ck}. We numerically calculate (\ref{eommix}) by starting from the positive frequency mode of the adiabatic vacuum $\left( \begin{array}{c} \tilde{u}_A \\ \tilde{u}_B \end{array} \right) = \left( \begin{array}{c} 1 \\ 0 \end{array} \right) e^{-i\int \omega_{\rm eff}dt}$ at remote past during inflation when all the relevant $k$-modes were well inside the Hubble horizon. The abundance of particles is calculated in terms of the Bogoliubov coefficients which relates mode functions of vacuum states at two different regimes, namely, de Sitter inflationary period and kination regime with power-law expansion $a(t) \sim t^{1/3}$. The energy density of gravitationally produced massive fermion is given by fitting (Fig. \ref{fig:rho})
\begin{equation}
	\rho \simeq 2\times10^{-3} e^{-4m\Delta t}m^2 H_{\rm inf}^2, \label{rho}
\end{equation}
where $\Delta t$ and $H_{\rm inf}$ denote the transition time scale from inflation to kination and the Hubble parameter during inflation, respectively. This is an order of magnitude larger than the conformally coupled massive scalar case \cite{Hashiba:2018iff}, however, the form of equation does not change.

\textit{Hierarchal right-handed neutrinos.}---We consider the following Lagrangian for the interaction of the right-handed Majorana neutrinos $N_i$ with masses $M_i$ ($M_1 \ll M_2 < M_3$),
\begin{equation}
	\mathcal{L}_N = M_i \bar{N}^c_i N_i + h_{i\alpha} N_i L_\alpha \phi^\dagger,
\end{equation}
where $L_\alpha, \phi$ and $h_{i\alpha}$ represent left-handed lepton doublets, the standard Higgs doublet and Yukawa coupling constant, respectively \cite{Fukugita:1986hr}. The decay rate of each right-handed neutrino $N_i$ is given by
\begin{equation}
	\Gamma_i = \frac{1}{4\pi} \sum_{\alpha} |\tilde{h}_{i\alpha}|^2 M_i, \label{Gamma}
\end{equation}
and the left-handed neutrino masses are derived by the seesaw mechanism \cite{Yanagida:1979as,GellMann:1980vs} as
\begin{equation}
	m_{\nu_\alpha} = \sum_i \left( \tilde{h}_{i\alpha} \right)^2 \cfrac{v^2}{2M_i}, \label{mnu}
\end{equation}
where $v \approx 246\ {\rm GeV}$, the mass matrix is diagonalized by the unitary Pontecorvo-Maki-Nakagawa-Sakata matrix $U_{i\alpha}$ \cite{Maki:1962mu} and $\tilde{h} \equiv h U^\ast$.
Hereafter, the right-handed neutrino masses are assumed to have a mass hierarchy $M_1 \lll M_2 \ll M_3$ as in the split seesaw mechanism \cite{Kusenko:2010ik}. According to neutrino oscillation observations such as the Super-Kamiokande \cite{Fukuda:1998mi}, KamLAND \cite{Araki:2004mb} and the MINOS \cite{Adamson:2011ig}, the heaviest and the second heaviest left-handed neutrino must have a mass around $0.05\ {\rm eV}$ and $0.01\ {\rm eV}$, respectively. Assuming the normal mass hierarchy, (\ref{mnu}) gives constraints on $\tilde{h}_{i\alpha}$
\begin{equation}
	\sum_i \left( \tilde{h}_{i\alpha} \right)^2 \frac{v^2}{2M_i} = m_{\nu_\alpha} \sim \begin{cases} 0.05\ {\rm eV} & (\alpha = 3) \\ 0.01\ {\rm eV} & (\alpha = 2) \end{cases}. \label{mnu2}
\end{equation}

We now discuss reheating, baryogenesis and generation of dark matter in turn, where each generation of right-handed neutrinos plays respective roles.

\textit{Reheating.}---First, the heaviest right-handed neutrino $N_3$ accounts for reheating. Since $\Delta t \simeq H_{\rm inf}^{-1}$ in our model, particles as heavy as $H_{\rm inf}$ are most efficiently produced by the gravitational particle creation as shown in (\ref{rho}). Therefore, the heaviest right-handed neutrino $N_3$ with mass $M_3 \sim H_{\rm inf}$ is produced most abundantly among all the particles and accounts for most of reheating process. The reheating temperature is derived from (\ref{rho}) and (\ref{Gamma}) as
\begin{align}
	T_{\rm RH} \simeq &6\times10^7\, \left( \frac{\sum_{\alpha} |\tilde{h}_{3\alpha}|^2}{10^{-12}} \right)^{-1/4}e^{-3M_3\Delta t} \nonumber \\
	&\times \left(\frac{M_3}{10^{13}{\rm GeV}}\right)^{5/4} \left(\frac{H_{\rm inf}}{10^{13}{\rm GeV}}\right)^{3/4} {\rm GeV}. \label{TRH}
\end{align}
In order to prevent the gravitationally produced graviton from disturbing CMB spectrum \cite{Aghanim:2018eyx}, $N_3$ should have a sufficiently long life time and behave as a non-relativistic matter for a while after its production. This condition is described by \cite{Hashiba:2018tbu}
\begin{equation}
	\left( \sum_{\alpha} |\tilde{h}_{3\alpha}|^2 \right)^{-1/3} e^{-4M_3\Delta t} \left(\frac{M_3}{H_{\rm inf}}\right)^{5/3} > 1.0 \times 10^2. \label{qgw}
\end{equation}
$e^{-4M_3\Delta t} \left( M_3/H_{\rm inf} \right)^{5/3}$ has a maximum value $4.4 \times 10^{-2}$ at $M_3 \simeq 0.42 H_{\rm inf}$ for $\Delta t = H_{\rm inf}^{-1}$ as a typical value. Hence, (\ref{qgw}) yields a constraint on the Yukawa coupling of $N_3$ as
\begin{equation}
	\sum_{\alpha} |\tilde{h}_{3\alpha}|^2 < 8.5 \times 10^{-11}. \label{h3aconst}
\end{equation}
If $\tilde{h}_{3\alpha}$ is as small as the Yukawa coupling of electron $\simeq 3 \times 10^{-6}$, then this condition is satisfied.

\textit{Baryogenesis.}---Next, we discuss baryogenesis through leptogenesis where $N_2$ provides CP violation. The one-loop correction shows that the decay of the right-handed neutrinos produces the lepton asymmetry with a non-vanishing CP violation phase and the Majorana mass term \cite{Fukugita:1986hr}. In our model, CP violation arises from the interference between $N_2$ and $N_3$. Since both of them are always out-of-equilibrium due to the relatively low reheating temperature, the net lepton number is supplied by the decay of both as
\begin{align}
	\frac{n_L}{s} &= \frac{\epsilon_3 n_3 a_\ast^{-3} + \epsilon_2 n_2 a_\ast^{-3}}{\frac{2\pi^2}{45}g_\ast \left[ \frac{30}{\pi^2 g_\ast} (M_3 n_3 a_\ast^{-4} + M_2 n_2 a_\ast^{-3}) \right]^{3/4}}, \label{leptoasym}
\end{align}
where $n_i$ and $\epsilon_i$ denote the number density of gravitationally produced $N_i$ when $N_3$ decays and the magnitude of the lepton asymmetry produced by the $N_i$ decay, respectively. Here, we assume that the scale factor has increased $a_\ast$-fold since $N_3$ decays until $N_2$ decays and they are non-relativistic before decaying \cite{Hashiba:2018iff}. $\epsilon_i$ is given by \cite{Flanz:1994yx,Covi:1996wh,Buchmuller:1997yu}
\begin{equation}
	\epsilon_i = -\frac{1}{8\pi} \frac{\sum_{\alpha \neq i} {\rm Im}\left[ \{ (hh^\dagger)_{i\alpha} \}^2 \right]}{(hh^\dagger)_{ii}} \left\{ f^V \left( x \right) + f^M \left( x \right) \right\}, \label{e1before}
\end{equation}
where $x = M_\alpha^2/M_i^2$, $f^V(x)$ and $f^M(x)$ represent the contribution from the one-loop vertex and self-energy corrections, respectively, and they are calculated as
\begin{align}
	f^V(x) &= \sqrt{x} \left[ -1 + (x + 1) \ln \left( \frac{x + 1}{x} \right) \right], \\
	f^M(x) &= \frac{\sqrt{x}}{x-1}.
\end{align}
Assuming a hierarchy $M_1/M_2 \ll M_2/M_3 \ll 1$, we obtain
\begin{align}
	\epsilon_2 &\simeq -\cfrac{1}{4\pi} \cfrac{1}{(\tilde{h}\tilde{h}^\dagger)_{22}} {\rm Im}\left[ \{ (\tilde{h}\tilde{h}^\dagger)_{23} \}^2 \right] \cfrac{M_2}{M_3}, \label{e2} \\
	\epsilon_3 &\simeq \cfrac{1}{4\pi} \cfrac{1}{(\tilde{h}\tilde{h}^\dagger)_{33}} {\rm Im}\left[ \{ (\tilde{h}\tilde{h}^\dagger)_{32} \}^2 \right] \cfrac{M_2}{M_3} \ln \cfrac{M_3}{M_2}. \label{e3}
\end{align}
$N_1$ must have a negligible contribution in order not to wash out the produced lepton asymmetry, which is the case because coupling of $N_1$ is sufficiently small to be nearly stable dark matter as discussed below. On the other hand, $a_\ast$ is given by
\begin{equation}
	a_\ast = \left( \frac{\sum_{\alpha} |\tilde{h}_{3\alpha}|^2 M_3}{\sum_{\beta} |\tilde{h}_{2\beta}|^2 M_2} \right)^{1/2}, \label{aast}
\end{equation}
assuming that the universe is radiation-dominant since $N_3$ decays until $N_2$ decays.

Equations (\ref{rho}), (\ref{Gamma}) and (\ref{e2}) -- (\ref{aast}) yield
\begin{widetext}
\begin{align}
	\epsilon_2 n_2 &\simeq 1.3 \times 10^{-5} \frac{{\rm Im}[ \{ (\tilde{h}\tilde{h}^\dagger)_{23} \}^2 ]}{(\tilde{h}\tilde{h}^\dagger)_{22}} \left( \sum_{\alpha} |\tilde{h}_{3\alpha}|^2 \right) M_2^2 H_{\rm inf}, \\
	\epsilon_3 n_3 &\simeq 1.3 \times 10^{-5} \frac{{\rm Im}[ \{ (\tilde{h}\tilde{h}^\dagger)_{32} \}^2 ]}{(\tilde{h}\tilde{h}^\dagger)_{33}} \left( \sum_{\alpha} |\tilde{h}_{3\alpha}|^2 \right) M_2 M_3 H_{\rm inf} e^{-4M_3 \Delta t} \ln\frac{M_3}{M_2},
\end{align}
\end{widetext}
and
\begin{align}
	M_2 n_2 a_\ast &\simeq 1.6 \times 10^{-4} \frac{(\sum_{\alpha} |\tilde{h}_{3\alpha}|^2)^{3/2}}{(\sum_{\beta} |\tilde{h}_{2\beta}|^2)^{1/2}} M_2^{3/2} M_3^{3/2} H_{\rm inf}, \\
	M_3 n_3 &\simeq 1.6 \times 10^{-4} e^{-4M_3 \Delta t} \left( \sum_{\alpha} |\tilde{h}_{3\alpha}|^2 \right) M_3^3 H_{\rm inf},
\end{align}
assuming $M_2 \Delta t \sim M_2/H_{\rm inf} \ll 1$. If $(\sum_{\alpha} |\tilde{h}_{3\alpha}|^2/\sum_{\beta} |\tilde{h}_{2\beta}|^2)^{1/2} \ll (M_3/M_2)^{3/2}$ is satisfied, then $M_2 \ll M_3$ yields $\epsilon_2 n_2 \ll \epsilon_3 n_3$ and $M_2 n_2 a_\ast \ll M_3 n_3$. Therefore, we may conclude that most of the resultant lepton asymmetry comes from the decay of $N_3$, while CP violation comes from the interference with $N_2$. The produced lepton asymmetry (\ref{leptoasym}) is given by
\begin{widetext}
\begin{equation}
	\frac{n_L}{s} \approx 3 \times 10^{-3} \frac{{\rm Im}[ \{ (\tilde{h}\tilde{h}^\dagger)_{32} \}^2 ]}{(\tilde{h}\tilde{h}^\dagger)_{33}} \left( e^{-M_3 \Delta t} \ln\frac{M_3}{M_2} \right) \left( \sum_{\alpha} |\tilde{h}_{3\alpha}|^2 \right)^{1/4} \frac{M_2}{M_3} \left( \frac{M_3}{H_{\rm inf}} \right)^{-1/4}.
\end{equation}
\end{widetext}
Since $M_3 \sim H_{\rm inf}$, we can let $e^{-M_3 \Delta t} \ln\frac{M_3}{M_2}$ and $M_3/H_{\rm inf}$ be an order of unity. This lepton asymmetry is finally converted into the baryon asymmetry by the sphaleron process \cite{Kuzmin:1985mm} as
\begin{equation}
	\frac{n_B}{s} = C \frac{n_L}{s}, \label{BL}
\end{equation}
where $C = 28/79$ in the standard model \cite{Khlebnikov:1988sr,Harvey:1990qw}. Therefore, taking the Yukawa coupling as almost the maximum allowed value (\ref{h3aconst}), we obtain
\begin{equation}
	{\rm Im}\biggl[ \Bigl( \sum_\alpha \tilde{h}_{3\alpha} \tilde{h}_{2\alpha}^\ast \Bigr)^2 \biggr] \sim 10^{-16} \frac{M_3}{M_2} \label{hh32}
\end{equation}
in order to realize $n_B/s = (8.65 \pm 0.06) \times 10^{-11}$ \cite{Cyburt:2015mya}. Assuming the normal mass hierarchy, $\tilde{h}_{31}$ and $\tilde{h}_{21}$ are much smaller than other $\tilde{h}_{3\alpha}$'s and $\tilde{h}_{2\alpha}$'s, respectively, and then the contribution from $\tilde{h}_{21}$ is negligible. Since (\ref{h3aconst}) yields $|\tilde{h}_{3\alpha}| \lesssim 10^{-5}$, (\ref{hh32}) means that
\begin{equation}
	\tilde{h}_{22} \ {\rm or} \ \tilde{h}_{23} \gtrsim 10^{-3}\sqrt{M_3/M_2}. \label{h2aconst}
\end{equation}
Since $\tilde{h}_{2\alpha}$ must satisfy (\ref{mnu2}), $M_2$ is constrained as $M_2 \gtrsim 10^{11}\ {\rm GeV}$. On the other hand, if $M_2$ and $M_3$ are almost degenerate (but their mass difference is still much larger than their decay widths $\Gamma_i$), then $f^M(x)$ dominates (\ref{e1before}) \cite{Akhmedov:1998qx,Fujii:2002jw,Pilaftsis:2003gt}. However, this is beyond the scope of this letter.

\textit{Dark matter.}---Finally, the lightest right-handed neutrino $N_1$ accounts for dark matter. $N_1$ must be so stable that its signal must be below the level detectable by the previous and ongoing X-ray observations such as Chandra, XMM-Newton, Suzaku and NuSTAR. These observations put stringent constraints on the mixing angle \cite{Perez:2016tcq}. For $M_1 \simeq 10\ {\rm keV}$, the mixing angle $\sin^2 2\theta$ should be smaller than $\sim 10^{-11}$. Therefore,
\begin{equation}
	\theta^2 \simeq \sum_\alpha |\tilde{h}_{1\alpha}|^2 \frac{v^2}{2M_1^2} < 10^{-11}
\end{equation}
and then the constraint on the Yukawa coupling is
\begin{equation}
	\sum_\alpha |\tilde{h}_{1\alpha}|^2 < 10^{-26}. \label{h1aconst}
\end{equation}
$N_1$ may also make the main contribution to the left-handed neutrino masses via the seesaw mechanism due to its extreme lightness, and then (\ref{mnu2}) gives
\begin{equation}
	|\tilde{h}_{1\alpha}|^2 < 10^{-21} \times \left( \frac{M_1}{10\ {\rm keV}} \right).
\end{equation}
This is included in the constraint (\ref{h1aconst}).

The original split seesaw \cite{Kusenko:2010ik} considers thermal production of $N_1$, but it requires a very high reheating temperature $\gtrsim 10^{11}\ {\rm GeV}$ which the gravitational reheating cannot achieve. Therefore, we have to use the gravitational particle creation also for production of dark matter $N_1$. Since its lightness makes the gravitational particle creation too inefficient to produce sufficient amount of $N_1$ to account for the dark matter, we introduce a non-minimal (or non-conformal) coupling between the scalar curvature $R$ and fermion $\psi$ such as $\frac{R}{\mu}\bar{\psi}\psi$, where $\mu$ is a constant with unit mass dimension. This term gives an effective mass $12H_{\rm inf}^2/\mu$ during inflation, which quickly vanishes after inflation. Incorporating this coupling in (\ref{Lpsi1}) and repeating the same numerical calculation, we find that the number density is given as $n \simeq 1.1\times10^{-1} H_{\rm inf}^5/\mu^2$ at the end of inflation for $\Delta t \approx H_{\rm inf}^{-1}$. Assuming that the universe evolves adiabatically after reheating with temperature (\ref{TRH}), we find that the right abundance of dark matter is realized if we take
\begin{widetext}
\begin{equation}
	\mu \approx 3 \times 10^{14}\ {\rm GeV} \times e^{\frac{3}{2}M_3 \Delta t} \left( \frac{M_1}{10\ {\rm keV}} \right)^{1/2} \left( \frac{M_3}{10^{13}\ {\rm GeV}} \right)^{-5/8} \left( \frac{H_{\rm inf}}{10^{13}\ {\rm GeV}} \right)^{13/8} \left( \frac{\sum_{\alpha} |\tilde{h}_{3\alpha}|^2}{8.5 \times 10^{-11}} \right)^{1/8}.
\end{equation}
\end{widetext}
This cutoff scale causes no undesirable non-perturbative effect since it is above the energy scale during inflation and below the Planck scale. Naively, we expect the same coupling to $N_2$ and $N_3$. For $N_3$, this coupling is negligible because $R/\mu$ is much smaller than $M_3$. For $N_2$, $R/\mu$ is comparable or even larger than $M_2$ and then the energy density of produced $N_2$ would be larger. However, this does not change the situation since the energy density of $N_2$ has no significant effect on neither the reheating nor the baryogenesis. Since the Yukawa coupling (\ref{h1aconst}) is quite small and $N_1$ is produced just after inflation, $N_1$ never reaches thermal equilibrium and its momentum is highly red-shifted, therefore, it becomes cold dark matter.

\textit{Discussion.}---Since quintessential inflation \cite{Geng:2017mic,Akrami:2017cir} gives late-time accelerated cosmic expansion, our model has the possibility of solving remaining big problems in cosmology --- reheating, baryogenesis, dark matter and dark energy. We have considered baryogenesis via the decay of the right-handed Majorana neutrinos produced by gravitational particle creation after quintessential inflation. In our scenario, both of the reheating process and baryogenesis are achieved by gravitational particle creation of three generations of right-handed Majorana neutrinos. This scenario can explain the reheating process via the decay of the heaviest right-handed neutrino with a mass $\sim H_{\rm inf} \simeq 10^{13}\ {\rm GeV}$, the present baryon asymmetry via the CP violation between the heaviest and the intermediate right-handed neutrino with a mass $\sim 10^{11}\ {\rm GeV}$ and the present dark matter abundance by the lightest right-handed neutrino with a mass $\sim10\ {\rm keV}$ with a higher dimensional operator $\frac{R}{\mu}\bar{\psi}\psi$ and a cutoff scale $\mu \sim 10^{15}\ {\rm GeV}$. The constrains on their Yukawa coupling are given in (\ref{h3aconst}), (\ref{h2aconst}) and (\ref{h1aconst}).

Although these conditions seem to require fine-tuning, the Randall-Sundrum (RS) brane-world scenario can explain these small couplings as well as the large mass hierarchy \cite{Randall:1999ee}. In the setup based on RS brane-world \cite{Kusenko:2010ik}, the effective four-dimensional (4D) mass $M_i$ and Yukawa coupling $\tilde{h}_{i\alpha}$ are rewritten as
\begin{align}
	M_i &= \kappa_i v_{\rm B-L} \frac{2m_i}{M(e^{2m_i l} - 1)} \\
	\tilde{h}_{i\alpha} &= \frac{\lambda_{i\alpha}}{\sqrt{M}} \sqrt{\frac{2m_i}{e^{2m_i l} - 1}} = \lambda_{i\alpha} \sqrt{\frac{M_i}{\kappa_i v_{\rm B-L}}},
\end{align}
where $\lambda_{i\alpha}, \kappa_i, m_i, M, l$ and $v_{\rm B-L}$ denote 5D Yukawa couplings, a numerical constant of order unity, a bulk mass, the 5D fundamental scale, the size of the extra dimension and $B-L$ breaking scale, respectively \cite{Kusenko:2010ik}. $M$ and $l$ are related to the 4D reduced Planck mass $M_G$ as $M_G^2 = M^3 l$. We take $\kappa_1 \sim 1, \kappa_2 \sim 0.1, \kappa_3 \sim 1, M \sim 5 \times 10^{17}\ {\rm GeV}, l^{-1} \sim 10^{16}\ {\rm GeV}$ and $v_{\rm B-L} \sim 10^{16}\ {\rm GeV}$ as reference values. In terms of 5D parameters, $M_3 \sim 10^{13}\ {\rm GeV}, M_2 \sim 10^{11}\ {\rm GeV}, M_1 \sim 10\ {\rm keV}$, (\ref{h3aconst}), (\ref{h2aconst}) and (\ref{h1aconst}) can be expressed as $m_3 \simeq 2.3 l^{-1}, m_1 \simeq 24 l^{-1}, \lambda_{3\alpha} < 3 \times 10^{-4}, \lambda_{22} \ {\rm or} \ \lambda_{23} \sim 1$ and $\lambda_{1\alpha} < 10^{-2}$, respectively. All of them are within the range of the Yukawa coupling of the SM fermions.

The lightest right-handed neutrino is within the range of NuSTAR observation \cite{Harrison:2013md} and then future observations may detect its signal. Since we take the almost maximum allowed value of $\tilde{h}_{3\alpha}$, future space-based gravitational wave observations may be able to detect quantum gravitational wave produced in the reheating era. Quintessential inflation itself will also be tested by future large-scale structure surveys as explained in \cite{Akrami:2017cir}.

\textit{Acknowledgments.}---We acknowledge useful comments of Kohei Kamada, Kazumi Kashiyama, Ayuki Kamada, Kazunori Nakayama and Yusuke Yamada. SH was supported by the Advanced Leading Graduate Course for Photon Science (ALPS). The work of JY was supported by JSPS KAKENHI, Grant JP15H02082 and Grant on Innovative Areas JP15H05888.

\bibliographystyle{apsrev4-1}
\bibliography{leptogenesis}

\end{document}